\documentclass[a4paper,12pt]{article}
\pdfoutput=1
\usepackage{bbm}
\usepackage{amsmath,times,amssymb,latexsym,graphics,braket,color,ascmac}
\usepackage{amsmath}
\usepackage[dvipdfm]{graphicx}
\usepackage{bm}
\usepackage{hyperref}
\usepackage{cite}

\newcommand{\be}{\begin{equation}}
\newcommand{\ee}{\end{equation}}
\newcommand{\nn}{\nonumber}

\newcommand{\lan}{\langle}
\newcommand{\ra}{\rangle}

\newcommand{\ex}[1]{\mathrm{e}^{#1}}

\newcommand{\pa}[1]{\left(#1 \right)}

\newcommand{\bb}[1]{\mathbb{#1}}

\newcommand{\ca}[1]{\mathcal{#1}}

\newcommand{\abs}[1]{\left|#1\right|}

\newcommand{\fr}{\frac}

\def\be{\begin{equation}}
\def\ee{\end{equation}}
\def\ba{\begin{eqnarray}}
\def\ea{\end{eqnarray}}

 \def\ba{{\bar{\alpha}}}

\def\ii{{\mathrm{i}}}
\def\tr{{\mathrm{tr}}}

\def\dd{{\mathrm{d}}}

\begin{document}
\begin{titlepage}
\begin{flushright}
{\small RIKEN-iTHEMS-Report-21}\\
{\small \today}
 \\
\end{flushright}

\begin{center}

\vspace{1cm}

\hspace{3mm}{\bf \Large Product of Random States and Spatial (Half-)Wormholes} \\[3pt] 

\vspace{1cm}

\renewcommand\thefootnote{\mbox{$\fnsymbol{footnote}$}}
Kanato Goto${}^{1}$, Yuya Kusuki${}^{1}$, Kotaro {Tamaoka}${}^{2}$, and Tomonori Ugajin${}^{3}$

\vspace{5mm}

${}^{1}${\small \sl RIKEN Interdisciplinary Theoretical and Mathematical Sciences (iTHEMS), \\Wako, Saitama 351-0198, Japan}\\
${}^{2}${\small \sl Department of Physics, College of Humanities and Sciences, Nihon University, \\Sakura-josui, Tokyo 156-8550, Japan}\\
${}^{3}${\small \sl Center for Gravitational Physics, 
Yukawa Institute for Theoretical Physics (YITP), \\Kyoto University, Kitashirakawa Oiwakecho, Sakyo-ku, Kyoto 606-8502, Japan}\\
\end{center}

\vspace{5mm}

\noindent
\abstract
We study how coarse-graining procedure of an underlying UV-complete quantum gravity gives rise to a connected geometry. It has been shown, quantum entanglement plays a key role in the emergence of such a geometric structure, namely a smooth Einstein-Rosen bridge. In this paper, we explore the possibility of the emergence of similar geometric structure from classical correlation, in the AdS/CFT setup. To this end, we consider a setup where we have two decoupled CFT Hilbert spaces, then choose a random typical state in one of the Hilbert spaces and the same state in the other. The total state in the fine-grained picture is of course a tensor product state, but averaging over the states sharing the same random coefficients creates a geometric connection for simple probes. Then, the apparent spatial wormhole causes a factorization puzzle. We argue that there is a spatial analog of half-wormholes, which resolves the puzzle in the similar way as the spacetime half-wormholes.

\end{titlepage}
\setcounter{footnote}{0}
\renewcommand\thefootnote{\mbox{\arabic{footnote}}}
\tableofcontents
\flushbottom
\section{Introduction and Summary}

Recent studies show wormholes capture important non-perturbative aspects of  gravity. 
For example, in order for the semi-classical description of a black hole to be consistent with the principle of quantum theory, we necessarily include replica wormholes into the relevant gravitational path integrals\cite{Penington:2019kki, Almheiri:2019qdq}(see  for example \cite{Goto:2020wnk,Colin-Ellerin:2020mva,Colin-Ellerin:2021jev,Balasubramanian:2020xqf,Balasubramanian:2020coy}). Wormholes also play a significant role in AdS/CFT correspondence\cite{Maldacena:1997re}. If we take a boundary with two disjoint components, $\mathcal{M}_{1} \cup  \mathcal{M}_{2}$, then its dual bulk geometry naturally contains wormholes connecting these two disjoint components. This apparently looks like a puzzle, because inclusion of such a bulk wormhole to the gravitational path integral, according to the bulk to boundary dictionary, implies the partition function of the boundary theory is not factorized to the contributions of each components 
$\mathcal{M}_{1}$ and $\mathcal{M}_{2}$.  This has been well-known as the factorization problem\cite{Maldacena:2004rf}. A resolution suggested was that the presence of wormholes in the bulk leads to an ensemble of field theories on the boundary, instead of a single field theory.  This  relation between wormholes in the bulk and the ensembles on the boundary was manifested in the correspondence between two-dimensional JT gravity and random  matrix theory\cite{Saad:2019lba}, also in a theory of topological gravity \cite{Marolf:2020xie}. These wormholes are all Euclidean, and are  sometimes referred to as spacetime wormholes.

In this paper, we consider another averaging operation, namely averaging over some set of states in a {\it fixed} theory, and discuss its dual gravity interpretation. 
Such state averaging naturally fits into the usual framework of AdS/CFT correspondence which relates a single quantum field theory on the boundary to a theory of gravity in the bulk.   Moreover, averaging over states is a natural operation from low energy effective theory point of view, therefore it  provides  an efficient  framework to study  quantum dynamics of black holes \cite{Page:1993df,Page:1993wv,Hayden:2007cs,Verlinde:2012cy}. 
Indeed, distinguishing two typical states requires measurement with enormously high energy, which is impossible for low energy observers. This is one of the fundamental principles of statistical mechanics. Parallel to this, in the bulk side, since we do not know the microscopic details of quantum theory of gravity, we are forced to employ a coarse-grained description which naturally leads to the aforementioned state averaging. Refer to \cite{Pollack:2020gfa, Belin:2020hea, Belin:2020jxr, Langhoff:2020jqa,Liu:2020jsv, Freivogel:2021ivu} for recent work in this direction.

A natural choice of such an ensemble is provided by a class of states called canonical thermal pure quantum (TPQ) states\cite{Sugiura:2011hm, Sugiura:2013pla},  
\begin{align}
\ket{\psi_\beta}=\dfrac{1}{\sqrt{Z(\beta)}}\sum_n \mathrm{e}^{-\frac{\beta}{2}E_n}c_n\ket{n}, 
\label{eq:Thestate}
\end{align}
where ${c_{n}}$ are random variables drawn from a particular distribution. Naively, we expect such TPQ states are realized naturally via time evolution in a chaotic system. There is a related concept such as eigenstate thermailzation hypothesis(ETH)\cite{Srednicki_1999, DAlessio:2015qtq, Mondaini_2017}. We review these ingredients in section \ref{sec:ens} and also discuss the relation between them from time to time.

One of the aims of this paper is to elucidate the relation between averaging over states in a fixed theory and spatial wormholes in its dual gravity description.
Spatial wormholes constitute another class of wormholes, and they are distinguished from the spacetime wormholes related to averaging over theories. A spacetime wormhole is a Euclidean manifold that connects two Euclidean boundaries, whereas a spatial wormhole can exist in a spacetime with a Lorentzian signature without violating causality. An example of a spatial wormhole is an Einstein-Rosen bridge of a two-sided black hole, which connects two asymptotic regions. 

\subsection*{Emergence of Einstein-Rosen bridge from classical correlations}

Our discussion is also motivated by the relation between quantum entanglement and the geometric structure of spacetime.
In the light of AdS/CFT correspondence, maximally extended AdS-Schwarzschild blackhole which has a spatial wormhole is dual to thermofield double (TFD) state\cite{Maldacena:2001kr, Israel:1976ur, Balasubramanian:1998de}, 
\begin{equation}\label{eq:TFD}
\begin{aligned}
\rho^{(\text{TFD})}_{LR}(\beta) &\equiv \ket{\text{TFD}_{\beta}} \bra{\text{TFD}_{\beta}},\\
\\
\ket{\text{TFD}_{\beta}}  &\equiv \dfrac{1}{\sqrt{Z(\beta)}}\sum_n\mathrm{e}^{-\frac{\beta}{2}E_n}\ket{n_Ln_R},
\end{aligned}
\end{equation}
where $\ket{n_{L,R}}$ correspond to energy eigenstates in the dual conformal field theories on the left and right boundaries, with the Hilbert spaces $\mathcal{H}_{R}$, $\mathcal{H}_{L}$\footnote{To be precise, we have to introduce  $\ket{n_{R}^\ast}\equiv\Theta\ket{n_{R}}$ instead of $\ket{n_{R}}$, where $\Theta$ is a CPT-like map which exchanges the left and right systems and reverses time. In this paper, we will be a bit sloppy about this point as long as the difference is not important. }. 
This is a concrete realization of the so-called  ER=EPR proposal\cite{Maldacena:2013xja}. This proposal suggests that the entanglement in the boundary theory is necessary to obtain a smooth spacetime connection on the gravity side. This suggestion brings up a naive but important question: {\it does this mean we can see spatial wormholes only when we have quantum entanglement?} This question has been also addressed in \cite{Verlinde:2020upt,Anegawa:2020lzw}.

In this paper, we argue that the low energy observer sharing a copy of TPQ states can see the illusion of spatial wormholes. To show this, we first  place a TPQ state $ \ket{\psi_\beta} $ to the right Hilbert space $\mathcal{H}_{R}$, and its copy to the left Hilbert space $\mathcal{H}_{L}$, so that the total state is the product of them $  \ket{\Psi_\beta}=\ket{\psi_{\beta L}}\otimes \ket{\psi_{\beta R}^*}\ $.  This is a factorized state, so there is no quantum correlation between left and right, in the fine-grained viewpoint. For example, if we compute a two-point function between the left and the right, it gets factorized, $\lan \Psi_\beta|\mathcal{O}_{L} \mathcal{O}_{R} |\Psi_\beta\ra = \bra{\psi_\beta} \mathcal{O}_{L} \ket{\psi_\beta} \bra{\psi_\beta^\ast} \mathcal{O}_{R} \ket{\psi_\beta^\ast}$.
However, if the observer can only perform low energy experiments, the observables such as the above two-point function  must be averaged over the random variables which appear in the definition of the TPQ states \eqref{eq:Thestate}.  We will show, this averaging generates the correlation between the left and the right, which mocks up quantum entanglement, at least for certain low energy operators. For more detail, please see section \ref{subsec:2pt}.

In the bulk point of view, the presence of the correlation between the left and the right due to the state averaging signals  the emergence of a spatial wormhole connecting two asymptotic regions. This is analogous to the fact that averaging over theories yields a spacetime wormhole in the bulk.
Operationally, such states can be prepared by using the local operation and classical communication (LOCC) where we cannot gain any quantum entanglement. More precisely, this is archived  by first preparing the set of random variables $\{c_{n}\}$, and distribute it to both the left and the right system. Thus, we can view it as a spatial wormhole as a consequence of {\it classical correlation} in the dual field theory. We discuss this in section \ref{subsec:ERC}.

This wormhole is a non-perturbative effect: it is not always a dominant saddle as spacetime wormholes in the spectral form factor\cite{Saad:2018bqo}. Although we can compute this contribution in a semi-classical way, we have to say this cannot have a macroscopic geometric dual in the usual sense. That is, if the correlation develops a macroscopic wormhole, the entanglement entropy should be proportional to $G_{N}^{-1}$, since this entropy is given by the cross section of the wormhole divided by $4G_N$. However, in section \ref{subsec:ee}, we will show that the entanglement entropy obtained in this way is at most $G_{N}^{-1}\mathrm{e}^{-1/G_{N}}$ even if we start from a coarse-grained state. This indicates that the present wormhole created by the classical correlation between the two systems and averaging over the states cannot be macroscopic. 

\subsection*{Factorization puzzle for low energy observers and ``spatial half-wormhole''}
\begin{figure}[t]
\centering
\includegraphics[scale=0.25]{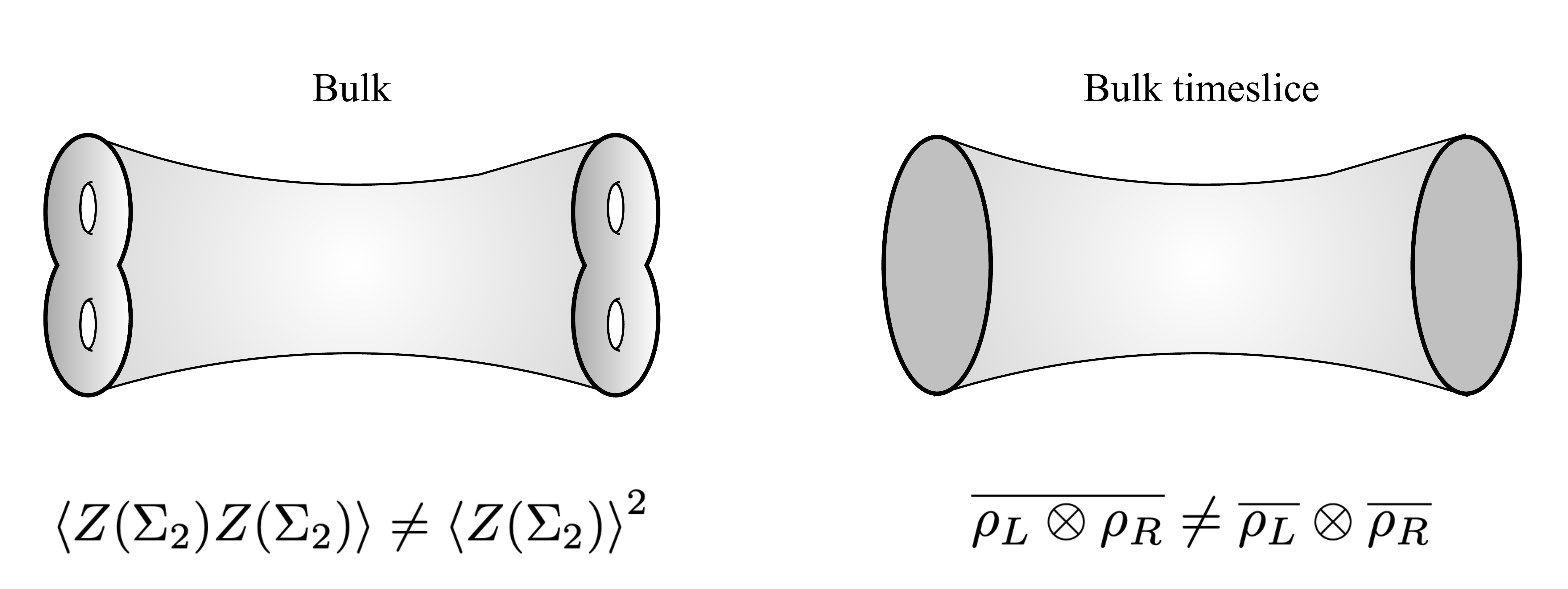}
\caption{Left: factorization problem based on spacetime wormholes. Right: a version of factorization problem based on spatial wormholes discussed in this paper. Throughout this paper, brackets $\braket{\cdots}$ (without suffix) represent the average over theories, while overlines $\overline{\cdots}$ represent the average over states.}\label{fig:wh}
\end{figure}

If a spatial wormhole exists in the bulk, a two-point function between the left and the right never gets 
factorized, as we can find a geodesic connecting these two boundaries (see Fig. \ref{fig:wh}). This is in tension with the fact that from the UV theory point of view, the total state is tensor factorized (see section \ref{subsec:fact}). We need to know the bulk mechanism which archives the factorization of the two-point function. 

In section \ref{sec:shw}, we show the factorization is provided by the analog of ``half-wormhole'' recently discussed to obtain a gravity dual of a single boundary theory\cite{Saad:2021rcu}, see also \cite{Mukhametzhanov:2021nea, Saad:2021uzi, Blommaert:2021gha,Garcia-Garcia:2021squ,Blommaert:2019wfy,Choudhury:2021nal}. To manifest the concept of half-wormhole, it is convent to introduce a copy $L^\prime R^\prime$ of the original system $LR$. 
In this doubled system $LRL^\prime R^\prime$, there are various new wormhole configurations exist, namely connecting the original system and its copy. These half-wormholes provide a crucial role to archive the factorization of the two-point functions connecting two boundaries, and therefore fine-grained viewpoint in gravity. 
We also discuss non-averaged systems and argue that non-self-averaging contributions in a single system can be identified with the spatial analog of half-wormhole. 

In section \ref{sec:discussion}, we discuss ambiguities from choice of ensembles and  interpretation of our result as generalization of bra-ket wormholes\cite{PhysRevD.34.2267, Chen:2020tes}. 

{\bf Note added:} While we were preparing this paper, we became aware of an interesting paper\cite{Freivogel:2021ivu} which independently discussed the similar results in section \ref{subsec:2pt} but from a different perspective, the quantum deviation. 

\section{Ensemble of states}\label{sec:ens}

In the next section, we will argue that  averaging over states on a bipartite system $\mathcal{H}_{L} \otimes \mathcal{H}_{R}$ leads to a correlation between two system, which signals the existence of spatial wormholes connecting these two boundaries. There are several notions of averaging over states or matrix elements that are closely related but different in detail.  We would like to figure out precisely which state averaging leads to a realistic wormhole structure in the bulk. The purpose of this section is to review  these notions and clarify the differences among them.

\subsection{Canonical TPQ states}\label{sec:TPQ}
 The first set of states over which we take the average is so-called (canonical) thermal pure quantum (TPQ) states \cite{Sugiura:2011hm, Sugiura:2013pla}\footnote{See also \cite{Nakagawa:2017yiw, Fujita:2018wtr} which compute the entanglement entropy using TPQ states. In particular, our normalization of the numerical coefficients $c_n$ is rather similar to them. \label{foot:tpq}}.  Such a  state  has the following form,
\begin{align}
\ket{\psi_\beta}=\dfrac{1}{\sqrt{Z(\beta)}}\sum_n \mathrm{e}^{-\frac{\beta}{2}E_n}c_n\ket{n}, \label{eq:cTPQ}
\end{align}
where $\ket{n}$ stands for an energy eigenstate and $\beta$ will be identified with the inverse temperature in the canonical ensemble. Here the numerical coefficients $c_n$ follow the Gaussian distribution. Thus when it is averaged over the distribution, we get $\overline{c_{m}} =0$ and
\begin{align}
    \overline{c_m^\ast c_n}&=\delta_{m,n},\\
    \overline{c_m^\ast c_n c_a^\ast c_b}&=\delta_{m,n}\delta_{a,b}+\delta_{m,b}\delta_{a,n},
\end{align}
for example. In general averages of the products of $c_{n}$s  are computed by the Wick's theorem. The overall factor is fixed such that norm of the state is unit after the averaging,
\begin{align}
\overline{\braket{\psi_\beta|\psi_\beta}}=\dfrac{1}{Z(\beta)}\sum_{m,n}\mathrm{e}^{-\frac{\beta}{2}(E_m+E_n)}\overline{c^\ast_mc_n}\braket{m|n}=1,
\end{align}
thus the normalization is identified with the thermal partition function with the inverse temperature $\beta$.
Also, the expectation values follow the thermal expectation values,
\begin{align}
\overline{\braket{\psi_\beta|\mathcal{O}|\psi_\beta}}=\dfrac{1}{Z(\beta)}\sum_n \mathrm{e}^{-\beta E_n}\braket{n|\mathcal{O}|n}\equiv\braket{\mathcal{O}}_\beta. \label{eq:TPQexp}
\end{align}
In this way, we can mimic thermal expectation values from the TPQ state. 
However, this does not imply that the TPQ state is effectively the thermal state.
If one would like advocate it, one should check that the fluctuation is small enough.

The upper bound on the fluctuation from the thermal expectation value can be estimated as
\footnote{
As noted in footnote \ref{foot:tpq}, our definition of the TPQ state is a little bit different from the original paper \cite{Sugiura:2011hm, Sugiura:2013pla}. 
Therefore, the estimation of the variance also has a small difference from their estimation. 
}
\begin{equation}\label{eq:flucTPQ}
\begin{aligned}
\overline{\pa{ \braket{\psi_\beta|\mathcal{O}|\psi_\beta} - \overline{\braket{\psi_\beta|\mathcal{O}|\psi_\beta}}^2}}
&=
\fr{1}{Z(\beta)^2} \sum_{n,m} \ex{-\beta(E_n + E_m)} \abs{\braket{n|\ca{O}|m}}^2\\
&\leq
\fr{1}{Z(\beta)^2} \sum_{n,m} \fr{\ex{-2\beta E_n}+\ex{-2\beta E_m}}{2}  \abs{\braket{n|\ca{O}|m}}^2\\
&\leq
\fr{1}{Z(\beta)^2} \sum_{n} \ex{-2\beta E_n}  \abs{\braket{n|\ca{O}|n}}^2.\\
\end{aligned}
\end{equation}
If we assume that the observables are low-degree polynomials ({\it i.e.}s, their degree $\leq m$, where $m=o(S)$),
\footnote{This type of operator is called as few-body operator. Note that the ETH is also based on a similar assumption.}
this fluctuation is exponentially suppressed as we expected since $Z(2\beta)/Z(\beta)^2\sim\mathrm{e}^{-\mathcal{O}(S)}$ where $S$ is the thermal entropy.
Therefore, the TPQ state can be effectively described by the thermal state in the averaging.

\subsection{Eigenstate Thermalization Hypothesis }\label{sec:ETH}

The thermalization from the averaging can also be obtained by an averaging over matrix elements in a similar way as the averaging over states.
We can impose the randomness on the matrix elements, instead of the coefficients of the state, like the form $\braket{n|\mathcal{O}|m} \simeq R_{nm} $ with the random variable $R_{nm}$,  which is nicely   captured by random matrix theory.
This type of structure is expected to be realized  in chaotic systems,
and summarized as  Eigenstate Thermalization Hypothesis (ETH) \cite{Srednicki_1999, DAlessio:2015qtq, Mondaini_2017},
\footnote{
In the following, since we will consider the canonical ensemble, so the coefficients of the random matrix would be meaningless. Nevertheless, we only use the randomness of the off-diagonal part, therefore, this subtlety does not matter in our paper. Here we just relate the matrix element averaging to the well-known fact in chaotic systems.
}
\footnote{
In particular, the systems of our interest, holographic 2D CFTs, show the ETH-like behavior as proved in \cite{Hikida:2018khg,Brehm:2018ipf,Romero-Bermudez:2018dim}.
}
\begin{equation}\label{eq:ETH}
\braket{n|\mathcal{O}|m} = \braket{\mathcal{O}(E)}_{th} \delta_{n,m} + \ex{-S(E)/2} g_O(E_n,E_m) R_{nm},
\end{equation}
where $\braket{\mathcal{O}(E)}_{th}$ is the thermal expectation value and $S(E)$ is the entropy at $E=\fr{E_n+E_m}{2}$.
The matrix $R_{nm}$ is a random Hermitian matrix with zero mean and unit variance. We will refer to averaging over the random matrix $R_{nm}$ as ETH averaging. The smooth function $g_\mathcal{O}(E_n,E_m)$ is of $\mathcal{O}(1)$ so that the off-diagonal part is suppressed by the dimension of the Hilbert space. 

The random average leads to
\begin{align}
\overline{R_{ij} R_{kl}} &= \delta_{i,l}\delta_{j,k},\\ \overline{R_{ij} R_{kl}R_{ab}R_{cd}} &= \overline{R_{ij} R_{kl}}\cdot \overline{R_{ab}R_{cd}}+\textrm{permutations}.
\end{align}
In other words, we can evaluate the product of the off-diagonal parts by using the Wick theorem.
This assumption is natural in our case since we now focus on the holographic CFT ({\it i.e.}, CFT dual to Einstein gravity), which is maximally chaotic system.

In a similar way as the TPQ state, we can see the thermalization by the ETH.
Let us consider a state,
\begin{equation}
\ket{\phi_\beta}=\fr{1}{\sqrt{Z(\beta)}} \sum_n \ex{-\fr{\beta}{2} E_n} \ket{n}.
\end{equation}
The expectation values are then,
\begin{equation}
\braket{\phi_\beta | \ca{O} | \phi_\beta} = \fr{1}{Z(\beta)} \sum_{n,m} \ex{-\fr{\beta}{2}\pa{E_n + E_m}} \braket{n|\ca{O}|m}.
\end{equation}
The ETH averaging for this expectation value leads to the same thermal expectation values,
\begin{equation}
\overline{\braket{\phi_\beta|\mathcal{O}|\phi_\beta}}=\dfrac{1}{Z(\beta)}\sum_n \mathrm{e}^{-\beta E_n}\braket{n|\mathcal{O}|n}. 
\end{equation}

The estimation of the fluctuation is similar to (\ref{eq:flucTPQ}) in the following way,
\begin{equation}\label{eq:flucETH}
\begin{aligned}
\overline{\pa{ \braket{\phi_\beta|\mathcal{O}|\phi_\beta} - \overline{\braket{\phi_\beta|\mathcal{O}|\phi_\beta}}^2}}
&=
\fr{1}{Z(\beta)^2} \sum_{n,m} \ex{-\beta(E_n + E_m)} \abs{\braket{n|\ca{O}|m}}^2\\
&\leq
\fr{1}{Z(\beta)^2} \sum_{n,m} \fr{\ex{-2\beta E_n}+\ex{-2\beta E_m}}{2}  \abs{\braket{n|\ca{O}|m}}^2\\
&\leq
\fr{1}{Z(\beta)^2} \sum_{n} \ex{-2\beta E_n}  \abs{\braket{n|\ca{O}|n}}^2\\
&\simeq
\fr{Z(2\beta)}{Z(\beta)^2}.
\end{aligned}
\end{equation}
In the last equation, we used $\braket{n|\ca{O}|n}=\mathcal{O}(1)$.
Therefore, the fluctuation from the thermal state is exponentially suppressed.

It would be interesting to note that even though our state averaging and ETH averaging look similar to each other,
they have a small difference if we evaluate higher-point correlators.
For example, the Wick contraction of $c_{n_1} c^*_{m_1} c_{n_2} c^*_{m_2} c_{n_3} c^*_{m_3} c_{n_4} c^*_{m_4}$ is different from that of
$R_{n_1 m_1} R_{n_2 m_2} R_{n_3 m_3} R_{n_4 m_4}$.
As we saw above, this difference does not appear in the one-point functions  and their variances. Therefore, it does not change the statement about thermalization.

\section{Spatial wormholes from product states}\label{sec:spatial}
In this section, we elucidate the relation between state averaging in  a chaotic theory  with a gravity dual and  spatial wormholes in  the bulk. To  study this, we consider  two disjoint  Hilbert spaces $\mathcal{H}_{L} \otimes \mathcal{H}_{R} $, then  prepare a tensor product  of the two identical states  $ \ket{\Psi_\beta}= \ket{\psi_{\beta L}}\otimes \ket{\psi_{\beta R}^\ast}$ on the total system. We study the correlation between two systems R and L which emerges after taking the average over a particular set of states. 
The simplest way to diagnose the correlation between two Hilbert spaces $\mathcal{H}_{L}, \mathcal{H}_{R}$ is  to compute a two-point function  $\braket{\Psi_\beta| \mathcal{O}_{R} \mathcal{O}_{L} |\Psi_\beta}$ of local operators.  Since the state $\ket{\Psi_\beta}$ on $\mathcal{H}_{L} \otimes \mathcal{H}_{R} $ is factorized,  the two-point function gets factorized as well. This corresponds to the measurement result of the observer with a fine-grained viewpoint, who knows everything about the state
$\ket{\Psi_\beta}$. On the contrary to this, if the observer can only access limited information, the measurement outcome is the two-point function averaged over sets of states $\overline{\braket{\Psi_\beta| \mathcal{O}_{R} \mathcal{O}_{L} |\Psi_\beta}}$,
because the observer can not distinguish the set of states.  
As we emphasized in the previous section, there are several notions of ``averaging over states'' which are natural for low energy measurements. We will discuss the difference of the resulting correlators based on different coarse-graining. 
\subsection{Correlation functions and spatial wormholes}\label{subsec:2pt}
\subsubsection*{Canonical TPQ analysis}

Let us first consider the set of TPQ states given by \eqref{eq:cTPQ},
\be
\ket{\Psi_\beta}=\ket{\psi_{\beta L}}\ket{\psi_{\beta R}^\ast}=\sum_{n,m}\ex{-\frac{\beta}{2}(E_n+E_m)}c_nc_m^*\ket{n_L}\ket{m_R}. \label{eq:tpq_double}
\ee
We are interested in the following  correlation function of the state,
\be
\braket{\Psi_\beta|\mathcal{O}_L\mathcal{O}_R|\Psi_\beta}=\dfrac{1}{Z(\beta)^2}\sum_{n,m,a,b}
\ex{-\frac{\beta}{2}(E_n+E_m+E_a+E_b)}
c_n^\ast c_m c_ac_b^\ast \braket{n|\mathcal{O}_L|a}\braket{m|\mathcal{O}_R|b}. \label{eq:LR1}
\ee

Note that the operators under consideration are not arbitrary as already mentioned in the previous section. When we discuss the holographic CFT, we assume $\mathcal{O}_L$ and $\mathcal{O}_R$ are so-called light operators. 

If we do not take the average, this correlation function can be written as a factorized form,
\be\label{eq:2pt0}
\braket{\Psi_\beta|\mathcal{O}_L\mathcal{O}_R|\Psi_\beta}=\braket{\psi_{L}|\mathcal{O}_L|\psi_{L}}\braket{\psi_{R}|\mathcal{O}_R|\psi_{R}},
\ee
where every contributions in the summand of \eqref{eq:LR1} are important to have the factorized answer. However  after taking the average, we obtain the following non-factorized result,
\begin{align}\label{eq:2pt}
&\overline{\braket{\Psi_\beta|\mathcal{O}_L\mathcal{O}_R|\Psi_\beta}}\nn\\
&=\braket{\mathcal{O}_L}_{\beta}\braket{\mathcal{O}_R}_{\beta}+\frac{Z(2\beta)}{Z(\beta)^2}\braket{\text{TFD}_{2\beta}|\mathcal{O}_L\mathcal{O}_R|\text{TFD}_{2\beta}}. 
\end{align}
Note that the second term is suppressed by $Z(2\beta)/Z(\beta)^2\sim\mathrm{e}^{-\mathcal{O}(S)}$ where $S$ is the thermal entropy. In other words, we eventually obtain a non-factorized contribution with order $e^{-\mathcal{O}(S)}$. As mentioned around \eqref{eq:TFD}, this term can be viewed as a contribution from the non-perturbative  spatial wormhole. 

If we consider a normalized combination, 
$
\overline{\braket{\Psi_\beta|\mathcal{O}_L\mathcal{O}_R|\Psi_\beta}}/\overline{\braket{\Psi_\beta|\Psi_\beta}}$,
we also have small negative corrections to the first and second terms in (\ref{eq:2pt}). Indeed such negative contributions sometime become important to keep unitarity (see recent work\cite{Stanford:2021bhl}, for example). Since this part does not affect our main claim, we will not track the denominator for the most part, and will come back to it in section \ref{sec:discussion}.

\subsubsection*{ETH}
A similar non-factorization can be found in the averaging over matrix elements. 
Let us consider the following state,
\begin{equation}
\ket{\Phi_\beta} =\ket{\phi_{\beta L}} \ket{\phi_{\beta R}}
=
\sum_{n,m}\ex{-\frac{\beta}{2}(E_n+E_m)}\ket{n_L}\ket{m_R}, \label{eq:ethd}
\end{equation}
where we assume that the energy eigenstate follows the ETH.
The non-averaged expectation value is obviously factorized as
\be
\braket{\Phi_\beta|\mathcal{O}_L\mathcal{O}_R|\Phi_\beta}=\braket{\phi_{L}|\mathcal{O}_L|\phi_{L}}\braket{\phi_{R}|\mathcal{O}_R|\phi_{R}},
\ee
however, in a similar way as the state averaging, we can see the non-factorization for the averaged expectation value,
\begin{align}
&\overline{\braket{\Phi_\beta|\mathcal{O}_L\mathcal{O}_R|\Phi_\beta}}\nn\\
&=\braket{\mathcal{O}_L}_{\beta}\braket{\mathcal{O}_R}_{\beta}+\frac{Z(2\beta)}{Z(\beta)^2}\braket{\text{TFD}_{2\beta}|\mathcal{O}_L\mathcal{O}_R|\text{TFD}_{2\beta}}. 
\label{eq:eth}
\end{align}
This sub-leading part is exponentially suppressed, which can be shown by the same estimation as (\ref{eq:2pt}).

\subsubsection*{Time-averaging}

Here we consider another averaging, time-averaging. 
We will again see that this example shows the correlation between two sides. However, the details of the expression of  the exponentially suppressed part are different from the above two examples. 


Let us consider a product state composed of two states,
\begin{align}\label{eq:product}
\ket{\Psi(t)}
&\equiv 
\ket{\psi_L(t)} \ket{\psi_R^*(t)}
\end{align}
where
\begin{align}
 \ket{\psi_L(t)}&=  {Z(\beta)}^{-\frac{1}{2}}\sum_n\ex{-\fr{\beta}{2}H_L} \ex{-iH_Lt}  \ket{n_L}, \\
 \ket{\psi_R^*(t)}&={Z(\beta)}^{-\frac{1}{2}}\sum_m\ex{-\fr{\beta}{2}H_R} \ex{iH_Rt}  \ket{m_R},
\end{align}
where $H_R \equiv 1\otimes H^T$ and $H_L \equiv H \otimes 1$.

The two-point function has the following spectral decomposition,
\begin{align}
\braket{\Psi(t)|\mathcal{O}_L\mathcal{O}_R|\Psi(t) } &= \sum_{n,m} e^{-\frac{\beta}{2}(E_{n}+E_{m})} \braket{m| \mathcal{O}_{L}|n}  e^{i(E_{m}-E_{n})t} \nonumber  \\ 
 &\times \sum_{k,l} e^{-\frac{\beta}{2}(E_{k}+E_{l})}\; \braket{k| \mathcal{O}_{R}|l}  e^{-i(E_{k}-E_{l})t} \nonumber  \\ 
 &\equiv \sum_{m,n,k,l} D_{(m,n,k,l)}(\beta) e^{i\left[(E_{m}-E_{n})-(E_{k}-E_{l}) \right]t}
 \label{eq:twopointtave}
\end{align}
Thus, if we integrate the correlation function over time, only the  part of the spectrum $\{E_{n} \}$ satisfying 
\be
(E_{m}-E_{n})-(E_{k}-E_{l}) =0 \label{eq:condition} 
\ee
contribute to the sum.  Apparently this contribution fixes one of the four sums in \eqref{eq:twopointtave}. However it is not obvious that there always be  a set of four energy eigenvalues satisfying the condition \eqref{eq:condition}.
For instance, it is widely believed that  in generic energy spectrum of a chaotic system, the condition is  satisfied only when  two of these eigenvalues agree,
\be
(E_{m}-E_{n})-(E_{k}-E_{l}) =0  \rightarrow 
\begin{cases}
E_{m} =E_{n},\; E_{k} =E_{l}, &{\rm  or} \\
E_{m} =E_{k},\; E_{n} =E_{l}.
\end{cases}
\ee
This is known as non-resonance condition. See
\cite{PhysRevA.30.504, Tasaki_1998, Srednicki:1995pt, Linden:2008awz, Smith:2012jea, Goldstein_2006} and references therein for the detail.

Once the condition is used, the time averaged  two-point function  is  given by
\begin{equation}\label{eq:time-ave}
    \overline{\braket{\psi_{LR}(t)|\mathcal{O}_L\mathcal{O}_R|\psi_{LR}(t)}}=\braket{\mathcal{O}_L}_{\beta}\braket{\mathcal{O}_R}_{\beta}+\frac{Z(2\beta)}{Z(\beta)^2}\tr \pa{\mathcal{O}_L\mathcal{O}_R \pa{\rho^{(\text{TFD})}_{LR}(2\beta) - \rho^{(\text{TMD})}_{LR}(2\beta)} },
\end{equation}
where we define the thermo mixed double(TMD) state\cite{Verlinde:2020upt, DelCampo:2019afl} as
\begin{equation}\label{eq:TMD}
\begin{aligned}
\rho^{(\text{TMD})}_{LR}(\beta) &\equiv\dfrac{1}{Z(\beta)}   \sum_n\mathrm{e}^{-\beta E_n}\ket{n_Ln_R} \bra{n_Ln_R}.
\end{aligned}
\end{equation}
See also recent studies\cite{Anegawa:2020lzw, Verlinde:2021kgt, Verlinde:2021jwu, Kudler-Flam:2021kfl} for holographic properties of the TMD state.
Note that in two-dimensional CFTs, we can approximate $\log Z \simeq \fr{\pi^2}{3}\fr{c}{\beta}$, $ S \simeq\fr{2}{3}\pi^2\fr{c}{\beta}$ and $E\simeq \fr{\pi^2}{3}\fr{c}{\beta^2}$ in the thermodinamic limit.
Therefore, the prefactor can be re-expressed by $Z(2\beta)/Z(\beta)^2 \simeq \ex{-\fr{3}{4}S}$.
When we apply the ETH on the second term, we obtain
\begin{equation}
\begin{aligned}
\frac{Z(2\beta)}{Z(\beta)^2}\tr \pa{\mathcal{O}_L\mathcal{O}_R \pa{\rho^{(\text{TFD})}_{LR}(2\beta) - \rho^{(\text{TMD})}_{LR}(2\beta)}}
&=
\frac{Z(2\beta)}{Z(\beta)^2}  \sum_{n\neq m} \fr{\braket{n|O_L\ex{-\beta H_L}|m} \braket{n|O_R \ex{-\beta H_R}|m}}{Z(2\beta)}
\\
&\leq
\frac{Z(2\beta)}{Z(\beta)^2} \fr{Z(\beta)^2 }{Z(2\beta)}  \underset{n\neq m}{\text{max}}\braket{n|O_L|m} \braket{n|O_R |m}\\
&= \mathcal{O}(\ex{-S(E)}).
\end{aligned}
\end{equation}
The second term is exponentially small compared to the first term.

Note that while the time-averaging also leads to the non-factorization, the sub-leading part is different from the state averaging (or the ETH) \eqref{eq:2pt} and \eqref{eq:eth}.
That is, the sub-leading part depends on a choice of ensemble.

\subsection{ER from classical correlations}\label{subsec:ERC}
Here we discuss how our results in the previous section can be interpreted as spatial wormholes and their origin as classical correlations.
In the previous section, we saw that the correlation function after averaging does not get factorized completely. These results have the following structure,
\be
 \overline{\braket{\psi_{LR}(t)|\mathcal{O}_L\mathcal{O}_R|\psi_{LR}(t)}} = 
 \braket{\mathcal{O}_L}_{th}\braket{\mathcal{O}_R}_{th}+ \mathrm{e}^{-\alpha S} D. \label{eq:nonfactorized}
\ee
In the above expression, $\alpha, D$ depend on the choice of averaging and stand for the non-factorized contribution. The correlation generated in this way is small, in the sense it is proportional to $\mathrm{e}^{-\mathcal{O}(S)}$, where $S$ is the thermal entropy associated to the inverse temperature $\beta$ in the state as in \eqref{eq:tpq_double} or \eqref{eq:ethd}. 

Through the AdS/CFT correspondence, we can interpret this result from the bulk AdS gravity point of view. Imagine that we compute this two-point correlation function from the gravity side  through the bulk to boundary dictionary, in terms of the bulk path integral in the  weak coupling limit $G_{N} \rightarrow 0$, and picking up the gravitational saddle points. The leading order in $G_{N}$ gives the factorized result, and it indicates that the leading bulk saddle  is disconnected and consists of two disjoint pieces. The second term in \eqref{eq:nonfactorized} which gives the correlation between two systems is coming from another saddle, because it is of order $\mathrm{e}^{-1/G_{N}}$. Therefore, it is natural to interpret that this second term is related to the regularized length of geodesic $L$ connecting two boundaries in this new saddle, {\it i.e.}, the second term is proportional to  $\mathrm{e}^{-L}\mathrm{e}^{-S}$.  The finiteness of $L$ implies two boundaries are connected in the new saddle. Said differently, the bulk geometry contains a spatial wormhole. In particular for TPQ state, the non-perturbative piece is given by the two-point function of the TFD state with the inverse temperature $\beta$.

We should also emphasize that different averaging procedures on the boundary lead to diffrerent values of the non-perturbative piece, although leading order in $G_{N} $ is always the same factorized result. This indicates that different averaging in the boundary CFT leads to different spatial wormholes in the bulk. In other words, the bulk spatial wormhole is highly depending on the specific averaging procedure on the boundary. 

It is also important to point out that the spatial wormholes we have been discussing so far are created by classical correlations. This is because the state can be prepared only by LOCCs which do not generate quantum entanglement. Let us take the TPQ state  $\ket{\Psi}=\ket{\psi_{\beta L}} \otimes \ket{\psi^\ast_{\beta R}}$, for example. To prepare the state, we fix the set of random variables $\{c_{n} \}$, then sending the set to Alice and Bob living in the left CFT and the right CFT by classical communication. Thus, correlation of the random coefficients between $\ket{\psi_{\beta L}}$ and $\ket{\psi_{\beta R}}$, which was the key to see the correlation, can be achieved in this way. Obviously, we cannot obtain the similar wormholes if we started from a product state $\ket{\psi_L}\ket{\phi_R}$, where $\ket{\psi_L}$ and $\ket{\phi_R}$ depend on independent random variables. 

\subsection{Entanglement entropy}\label{subsec:ee}
Since the state discussed so far is factorized, the two systems are not entangled at all, in the fine-grained picture.
On the other hand, averaging over the coefficients $\{c_{n}\}$ generates TFD like correlation between two systems and that generates the Einstein-Rosen bridge-like geometry from the gravity point of view. Therefore, it would be natural to ask whether the resulting geometry can be viewed as a macroscopic wormhole, and if it cannot, how it is different from the ordinary classical Einstein-Rosen bridge.

Motivated by this, in this section we would like to evaluate the entanglement entropy of our coarse-grained states. We remark that the rule to evaluate the entropy here is, first coarse-graining the state by taking the random averages for all coefficients $\{c_{n}\}$, and evaluate the entanglement entropy for such a coarse-grained state. In the later discussion section \ref{sec:discussion}, we will also make some remarks on how the entanglement entropy changes if we compute it in reverse order; writing the relevant quantities in the fine-grained picture, {\it i.e.} keeping all coefficients $\{c_{n}\}$, then taking the random average at the very end of the calculation.
  
We will evaluate the entanglement entropy for the following coarse-grained state, which can be written as a linear combination of the product of thermal states and the TFD states obtained by averaging  \eqref{eq:tpq_double},
\begin{align}
\overline{\rho_{LR}}&=\overline{|{\Psi_\beta}\rangle\langle{\Psi_\beta}|}/\overline{\braket{\Psi_\beta|\Psi_\beta}}\label{eq:cgstate}\\
&=p_1 \rho_{\beta L}\otimes\rho_{\beta R}+p_2\ket{\text{TFD}_{2\beta}}\bra{\text{TFD}_{2\beta}},  \label{eq:CS}
\end{align}
where
\begin{align}
p_1&=\dfrac{Z(\beta)^2}{Z(\beta)^2+Z(2\beta)}\sim {\cal O}(1),\\
p_2&=\dfrac{Z(2\beta)}{Z(\beta)^2+Z(2\beta)}\sim \mathrm{e}^{-{\cal O}(S)}.
\end{align}

We would like to estimate size of ``cross-section'' and length of ``throat'' of our wormhole via the entanglement entropy. In doing so, it is useful to remind that the von-Neumann entropy for a state $\rho=\sum_ip_i\rho_i$ satisfies the following inequality (see \cite{Nielsen_2009}, for example),
\be
\sum_ip_iS(\rho_i)\leq S(\rho)\leq \sum_ip_iS(\rho_i)+\sum_i(-p_i\log p_i). 
\ee
Therefore, the $S(\rho)$ is at most
\be
\label{entropyatmost}
S(\rho)\overset{\text{max}}{=}\sum_ip_iS(\rho_i)+\sum_i(-p_i\log p_i).
\ee
Thus, the maximal value of the entropy can be estimated as a sum of the entanglement entropy for each $\rho_i$. Each $\rho_i$ can be considered to be dual to the semi-classical geometry, thus the entropy for each $\rho_i$  gives the semi-classical contribution ${\cal O}(G_N^{-1})$. However, suggested by (\ref{entropyatmost}), the contribution from each entropy is suppressed by $p_i$.

First, we discuss its cross-section that is estimated by entanglement entropy between $L$ and $R$. The inequalities say that since each thermal state has an entropy of order $G_N^{-1}$, the resulting entanglement entropy between $L$ and $R$ is also of order $G_N^{-1}$. Therefore, at this point, we cannot conclude that our wormhole is non-perturbatively small, although it is clear that most of the contribution comes from the product state. 

On the other hand, we can conclude length of the wormhole throat is non-perturbatively small even for the coarse-grained state. To be explicit, let us consider two-dimensional holographic CFT${}_{L,R}$ on a compact space ({\it i.e.} circle) and trace over some part for each CFT, say $\bar{A}_{L,R}$. The resulting reduced density matrix for \eqref{eq:CS} is 
\be \label{eq:RDM}
\overline{\rho_{A_LA_R}}=p_1 \rho_{1 A_LA_R}+p_2\rho_{2 A_LA_R},
\ee
where
\begin{align}
\rho_{1 A_LA_R}&=\tr_{\bar{A}_{L}}\rho_{\beta L}\otimes\tr_{\bar{A}_{R}}\rho_{\beta R},\\
\rho_{2 A_LA_R}&=\tr_{\bar{A}_{L}\bar{A}_{R}}\ket{\text{TFD}_{2\beta}}\bra{\text{TFD}_{2\beta}}.
\end{align}
If we consider a time evolution at early time as \cite{Hartman:2013qma}, the growth of wormhole, coming only from the second term of the right side of \eqref{eq:RDM}, is captured by $S(\rho_{2 A_LA_R})=\frac{\alpha}{4G_N} t$, where $\alpha$ is an order $G_N^0$ coefficient. This is suppressed by $\mathrm{e}^{-\mathcal{O}(S)}$. This implies that the contribution from the TPQ wormhole is at most order $\mathrm{e}^{-\frac{1}{G_N}}G_N^{-1}$, hence not like the ordinary macroscopic wormhole.

\subsection{A factorization puzzle}\label{subsec:fact}
The above discussion can be interpreted as a version of factorization problem based on spatial wormholes. Namely, although we started from a factorized state (product state), the averaging gives rise to a non-product contribution which can be viewed as a contribution from the spatial wormhole. This becomes sharp especially if we consider a charged operator $\mathcal{O}_{q}$ whose thermal expectation value is vanishing (see also related work based on spacetime wormholes\cite{Belin:2020jxr}). 
On the one hand, we have
\begin{align}
    \overline{\braket{\psi_\beta|\mathcal{O}_{q}|\psi_\beta}}=\braket{\mathcal{O}_{q}}_\beta=0.
\end{align}
On the other hand, we can prepare a particular pair of charged operators $\mathcal{O}_q$ and $\mathcal{O}_{q^\prime}$ such that
\begin{align}
    \overline{\bra{\Psi_\beta}\mathcal{O}_{q L}\mathcal{O}_{q^\prime R}\ket{\Psi_\beta}}&=\overline{\bra{\psi_{\beta L}}\mathcal{O}_{q L}\ket{\psi_{\beta L}}\bra{\psi_{\beta R}^\ast}\mathcal{O}_{q^\prime R}\ket{\psi_{\beta R}^\ast}}\nn\\
    &\propto\braket{\text{TFD}_\beta|\mathcal{O}_{q L}\mathcal{O}_{q^\prime R}|\text{TFD}_\beta}\neq0.
    \label{eq:charged2pt}
\end{align}

The similar situation has been discussed by Harlow in \cite{Harlow:2015lma}. That is, having a Wilson line penetrating  a spatial  wormhole seems at odds with the structure of a microscopic (physical) Hilbert space which should be factorized, but at the same time it is necessary to keep the gauge invariance. The proposed resolution is that if there are heavy charged degrees of freedom in the bulk UV Hilbert space, the Wilson line can be split into two pieces at the bifurcation surface. In particular, from the IR point of view, we cannot distinguish whether or not the Wilson line is terminated on the bifurcation surface.  

Although our present example \eqref{eq:charged2pt}  only involves a specific state, therefore it is not about  the problem of factorization of Hilbert space itself,  what we have demonstrated so far is quite reminiscent of it. In our example, the source of the puzzle is clear. When we coarse-grained, we neglected a bunch of off-diagonal elements (or non-self-averaging elements) in the original product states. In other words, in the coarse-grained picture, there are missing degrees of freedom which originally helped the state factorization. From this perspective, the missing degrees of freedom may be identified with so-called edge modes in gauge theory and gravity\cite{Donnelly:2011hn, Donnelly:2014fua, Donnelly:2016auv, Takayanagi:2019tvn}.

Such degrees of freedom can revive in the higher moment of correlation functions as these are more fine-grained observables. 
In the next section, we will link the missing degrees of freedom to spatial analog of ``half-wormholes''.

\section{Higher moments and spatial half-wormholes}\label{sec:shw}
In this section, we argue that a product state can have a similar structure to the ``half-wormhole'', which is first proposed in \cite{Saad:2021rcu}. 

To summarize briefly, we will see how TPQ states can split into self-averaging and non-self-averaging parts. Then, we will show a square of the non-self-averaging part indeed gives a spatial wormhole which is self-averaging. We will also discuss the entire calculations follow an analog of the ``wormhole Wick theorem'' and hold a similar structure as (spacetime) half-wormholes. 

\subsection{Review of Half-Wormhole}
Here we will briefly summarize the results in \cite{Saad:2021rcu} and introduce the concept of half wormhole.

To see how the factorization problem is resolved, an interesting toy model was provided by Saad, Shenker, Stanford and Yao (called as SSSY model).
They considered  following  finite dimensional Grassman integral,
\begin{equation}\label{eq:SYK}
z=\int \dd^N\psi \  \ex{ \ii^{\fr{q}{2}}\sum_{1 \leq i_1 \leq \cdots \leq i_q \leq N} J_{i_1,i_2, \cdots , i_q} \psi_{i_1,i_2, \cdots , i_q} },
\ \ \ \ \ \ \ 
\psi_{i_1,i_2, \cdots , i_q} \equiv \psi_{i_1} \psi_{i_2} \cdots \psi_{i_q},
\end{equation}
 which is structurally similar to   the SYK partition function, except the fact that  the time direction is reduced to a point. In the same way as the standard SYK,  the average is taken over the $J$ tensor, which follows the Gaussian distribution.
In this model, the average of $z$ vanishes, $\braket{z}=0$, therefore,
the wormhole contribution can be dominant in the average $\braket{z_L z_R}$, due to the absence of the disconnected part.
They showed that those saddle points can be really interpreted as ``wormhole saddle point''.
Their approach is as follows.
The starting point is the expression,
\begin{equation}
z_L z_R 
=
\int \dd^N\psi^L \dd^N\psi^R \  \ex{ \ii^{\fr{q}{2}}\sum_{1 \leq i_1 \leq \cdots \leq i_q \leq N} J_{i_1, \cdots , i_q} \pa{\psi^L_{i_1, \cdots , i_q}+ \psi^R_{i_1, \cdots , i_q} }}.
\end{equation}
To study the connection between two systems, it is useful to introduce the following identity,
\begin{equation}
\begin{aligned}
1&=\int_{\bb{R}} \dd G_{LR} \ \delta\pa{G_{LR}-\fr{1}{N}\sum_{i=q}^N \psi_i^L \psi_i^R} \ex{\fr{N}{q}\pa{G_{LR}^q-\pa{\fr{1}{N} \sum_{i=q}^N \psi_i^L \psi_i^R}^q } }  \\
&=
\int_{\bb{R}} \dd G_{LR}  
\int_{\ii \bb{R}} \fr{\dd \Sigma_{LR}}{2\pi i / N}
\ex{-\Sigma_{LR}\pa{NG_{LR}-\sum_{i=q}^N \psi_i^L \psi_i^R}}
\ex{\fr{N}{q}\pa{G_{LR}^q-\pa{\fr{1}{N} \sum_{i=q}^N \psi_i^L \psi_i^R}^q } },
\end{aligned}
\end{equation}
where $G_{LR}$ and $\Sigma_{LR}$ are called  ``collective fields''.
If we evaluate the averaged partition function $\braket{z_L z_R}$ by inserting this identity,
we obtain
\begin{equation}
\braket{z_L z_R} = 
\int_{\bb{R}} \dd g
\int_{\ii \bb{R}} \fr{\dd \sigma }{2\pi i / N}
\ex{ N\pa{\log \pa{\ii \ex{-\fr{\ii \pi}{q} }\sigma}   - i\sigma g - \fr{1}{q} g^q  }  },
\end{equation}
where we introduced  "rotated"  collective fields,
\begin{equation}
\Sigma_{LR} = \ii \ex{-\fr{\ii \pi}{q} }\sigma,
\ \ \ \ \ \ \ 
G_{LR}=\ex{\fr{\ii \pi}{q}}g.
\end{equation}
From this expression, we can find that the saddles in the large $N$ limit are located on a unit circle $\abs{\sigma}=1$.
Roughly speaking, these collective fields represent a correlation between two systems $L$ and $R$ (that is, the saddle at $\sigma=0$ corresponds to the disconnected geometry without  wormhole  and the saddles away from the origin correspond to connected, or ``wormhole'', saddles), because in the full SYK model $G_{LR}$ is interpreted as the propagator between $L$ and $R$.

In \cite{Saad:2021rcu},  they were mainly interested in  the  ``non-averaged'' partition function $z_L z_R$, and its saddle points. Again  by writing the partition function
in terms of an integral over the collective fields,  they found two types of saddle points.
One of them  is the wormhole saddles  distributed on the unit circle $|\sigma| =1$, which also appear in $\braket{z_L z_R}$. This  implies that 
these saddles  are also saddles of  the non-averaged partition function, $z_L z_R \supset \braket{z_L z_R}$.
Furthermore, there is  another type of saddle point localized at $\sigma=0$, whose contrubution is denoted by $\Phi(0)$.
They call this saddle as ``half-wormhole''.
Since the half-wormhole saddle strongly depends on the coupling $J_{1_{1} \cdots i_{q}}$, the contribution of this saddle disappears in the averaged partition function  $\braket{z_L z_R}$.
For this reason, this saddle is sometimes referred to as non self-averaging point.
In contrast to this, an ordinary wormhole saddle is called  a self-averaging point.
The existence of  half-wormhole saddle is manifested in $\braket{z_L z_R z_{L'} z_{R'}}$,
where $L,R$ are the original systems of our interest and $L',R'$ are the auxiliary systems.
In the large $N$ limit, we can find the decomposition by the Wick contraction,
\begin{equation}\label{eq:z4}
\braket{z_L z_R z_{L'} z_{R'}} = \braket{z_L z_R} \braket{z_{L'} z_{R'}} +  \braket{z_L z_{L'}} \braket{z_R z_{R'}}  +  \braket{z_L z_{R'}} \braket{z_{L'} z_R}.
\end{equation}
The saddle points on $\abs{\sigma}=1$ only reproduce the first term.
It implies that the non-averaged partition function $z_L z_R$ should have other saddles.

They proposed that  $z_L z_R$ can be well-approximated by wormhole and half-wormhole saddles, $\braket{z^2} + \Phi(0)  (\equiv \mathbf{z}^2)$.
For this to be true, the error $\braket{ (z^2 - \mathbf{z}^2)^2  }$ should be small.
In \cite{Saad:2021rcu}, they checked that this error is indeed suppressed, which ensures the proposal that these two saddles are enough to approximate $z_L z_R$,
\begin{equation}
z^2 \simeq \braket{z^2} + \Phi(0).
\end{equation}

Note that, from  (\ref{eq:z4}), $z^2$ is expected to have the following suggestive form
\begin{equation}\label{eq:normal}
z^2 \simeq \braket{z^2} + :z^2: ,
\end{equation}
where we define $:z^2:$ as the normal ordered product, which follows the Wick theorem, except that we exclude a self-contraction  (for example, $\braket{:z_A z_B: :z_C z_D:} = \braket{z_A z_C}\braket{z_B z_D} + \braket{z_A z_D}\braket{z_B z_C}  $ ) .
This structure can also be found in a direct calculation of $z^2$ itself \cite{Mukhametzhanov:2021nea}.
Note also that the name ``half-wormhole'' comes from the reason that if we focus on the original systems $L,R$ in (\ref{eq:z4}),
the second and third term corresponds a remnant of the wormhole to the auxiliary systems $L', R'$.
In the next section, we will show that a product state can have a similar structure,
a decomposition into the wormhole and half-wormhole contributions. 

\subsection{Spatial half-wormholes}\label{subsec:shw}
In this section, we would like to show that  an analog of the half-wormholes also appears  in the system with  state  averaging discussed in the previous sections.
That is, (1) we have a non-self-averaging contribution in a single TPQ state that, given more than one copy, produces new self-averaging contributions (spatial wormholes). (2) Furthermore, it satisfies the ``wormhole Wick theorem''.

\subsubsection{Wormhole Wick theorem}\label{subsubsec:WWT}

As a warm-up exercise, let us revisit a single TPQ state, 
\begin{align}
\rho(c,\beta)\equiv\ket{\psi}\hspace{-1mm}\bra{\psi}=\sum_{nm}\mathrm{e}^{-\frac{\beta}{2}E_n}\mathrm{e}^{-\frac{\beta}{2}E_m} c_n c_m^* \ket{n}\hspace{-1mm}\bra{m}.
\end{align} 
The goal is to split it into the self-averaging part and non-self-averaging part. To this end, we write
\begin{align}\label{eq:splitc}
    c_nc_m^\ast=\overline{c_nc_m^\ast}+:c_nc_m^\ast:.
\end{align}
Here we defined  ``normal ordering'' of the random valuables  $:c_nc_m^\ast:\equiv c_nc_m^\ast-\overline{c_nc_m^\ast}$.
In fact, this normal ordering follows the standard rule, {\it i.e.},  summing over all possible Wick contractions except for the  self-interaction. For example,
\begin{equation}
    \overline{:c_n c_m^*:} = 0.
\end{equation}
\begin{equation}\label{eq:RR}
    \overline{:c_n c_m^*::c_p c_q^*:}
    =\overline{c_n c_q^*} \cdot \overline{c_m c_p^*}
    =\delta_{n,q} \delta_{m,p}.
\end{equation}
It implies that $:c_n c_m^*:$ can be thought of as a random variable with zero mean and unit variance.
For this reason, we denote
\begin{equation}
    R_{nm} \equiv :c_n c_m^*:. \label{eq:RTPQ}
\end{equation}

By using this expression, we can rewrite TPQ doubled state as
\begin{align}
\rho_{LR}(c,\beta)=\ket{\psi_L}\hspace{-1mm}\bra{\psi_L}\otimes\ket{\psi_R^*}\hspace{-1mm}\bra{\psi_R^*}=\sum_{nm}\mathrm{e}^{-\frac{\beta}{2}(E_n+E_m+E_a+E_b)}c_n c_m^* c_a^* c_b \ket{n_La_R}\hspace{-1mm}\bra{m_Lb_R}. 
\end{align}
To see the structure of it, again let us compute the average of the two point function,
\begin{align}
\mathrm{Tr}(\rho_{LR}(c,\beta)\mathcal{O}_L\mathcal{O}_R). \label{eq:tpqd}
\end{align}
We would like to reproduce the result \eqref{eq:2pt} by using \eqref{eq:splitc}-\eqref{eq:RR}.  To this end, we need to compute
\begin{align}
    \overline{c_n c_m^* c_a c_b^*}=\overline{c_nc_m^\ast}\,\cdot\,\overline{c_ac_b^\ast}+\overline{:c_nc_m^\ast::c_ac_b^\ast:} \label{eq:CCbar}
\end{align}
The first term gives a ``self-loop'' in the summand,
\begin{align}
    \overline{(\text{the first term in \eqref{eq:CCbar}})}=\delta_{n,m}\delta_{a,b},
\end{align}
and results in a product of thermal expectation values,
$\braket{\mathcal{O}_L}_\beta\braket{\mathcal{O}_R}_\beta$.
The second term is just given by (\ref{eq:RR}).
Thus, for few-body operators, the averaged density matrix is given by
\begin{equation}
\overline{\rho_{LR}(c,\beta)} = \rho_{L}(\beta) \otimes \rho_{R}(\beta) + \frac{Z(2\beta)}{Z(\beta)^2} \rho_{LR}^{\text{(TFD)}}(2\beta),
\end{equation}
where we defined $\rho_{L,R}(\beta)$ as the canonical ensemble,
\begin{equation}
\rho_{L,R}(\beta) \equiv  \dfrac{1}{Z(\beta)}\sum_n \ex{-\beta H} \ket{n_{L,R}} \bra{n_{L,R}}.
\end{equation}
The second term explains the spatial wormhole contributions as we have already discussed in section \ref{sec:spatial}. In this way, as more copies of the state are introduced, more wormhole contributions that connect originally disconnected systems appear. 

This re-derivation manifests the following: each TPQ state $\rho_{L,R}(c,\beta)$ consists of a self-averaging part $\overline{c_n c_m^\ast}=\delta_{n,m}$ and a non-self-averaging part $:c_nc_m^\ast:=R_{nm}$.
In particular, the ``square of $R_{nm}$'' gives a standard spatial wormhole (an Einstein Rosen bridge), a new self-averaging contribution according to the Wick theorem. 

\newsavebox{\boxshwl}
\sbox{\boxshwl}{\includegraphics[width=65pt]{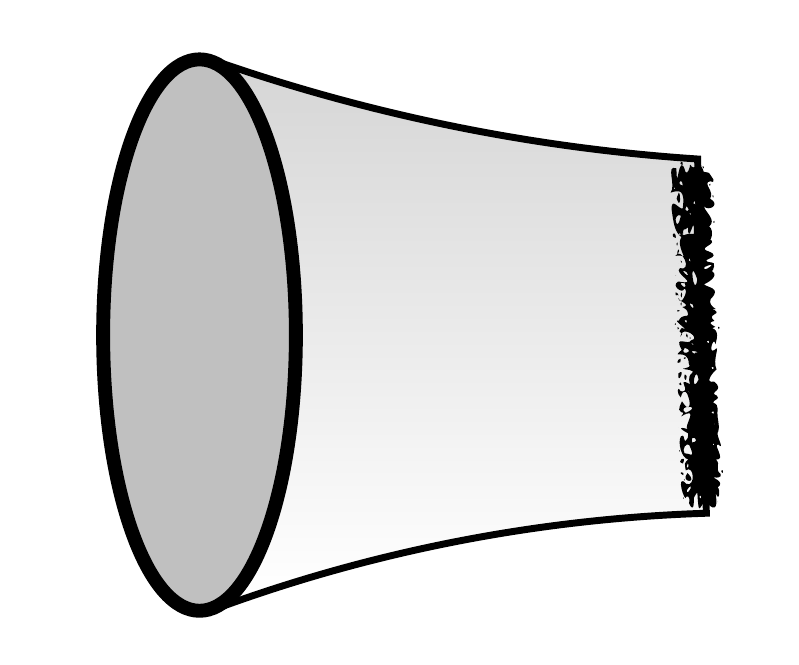}}
\newlength{\shwl}
\settowidth{\shwl}{\usebox{\boxshwl}} 

\newsavebox{\boxshwr}
\sbox{\boxshwr}{\includegraphics[width=65pt]{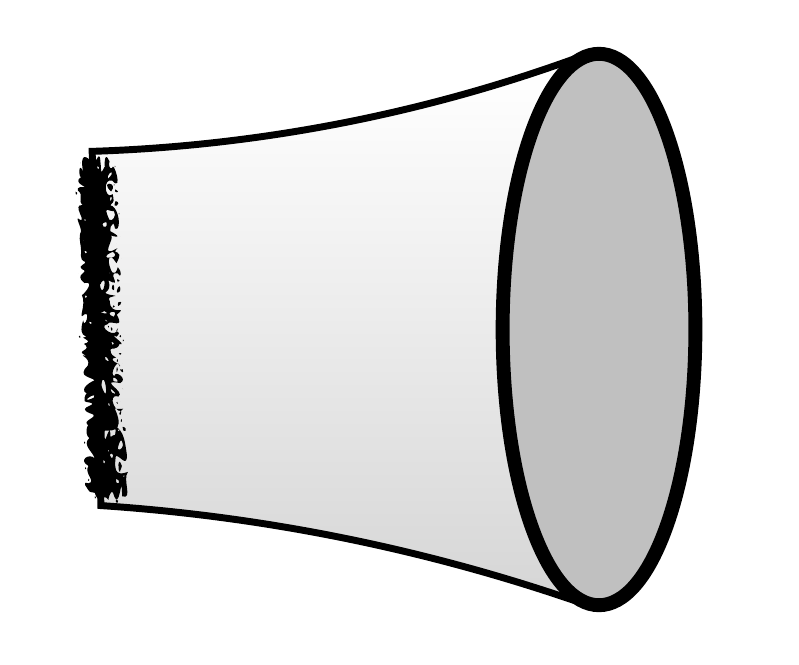}}
\newlength{\shwr}
\settowidth{\shwr}{\usebox{\boxshwr}} 

\newsavebox{\boxwh}
\sbox{\boxwh}{\includegraphics[width=95pt]{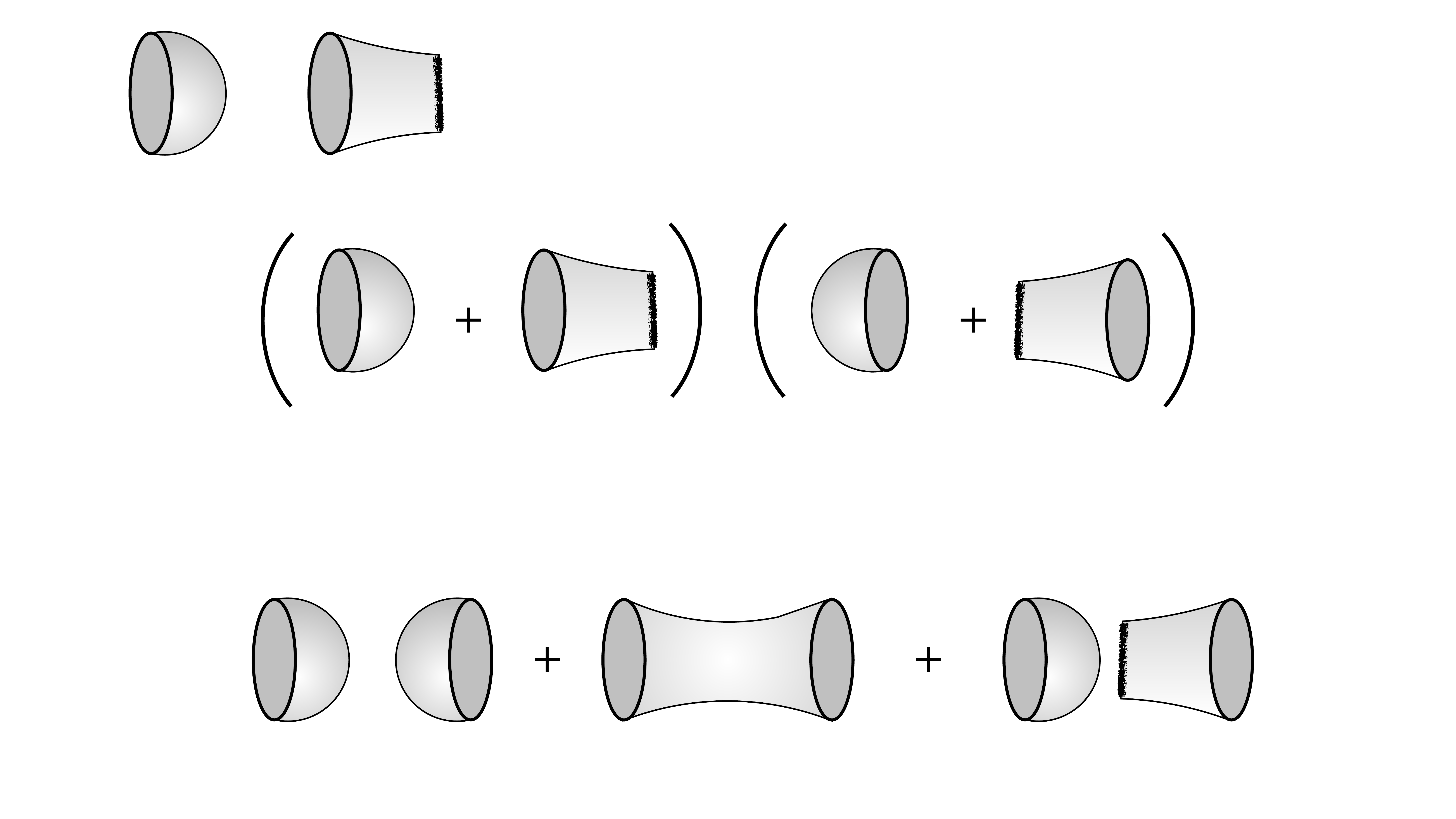}}
\newlength{\wh}
\settowidth{\wh}{\usebox{\boxwh}}

The argument so far motivates us to write our TPQ state as
\begin{equation}
\rho(c,\beta)= \rho(\beta)+:\rho(c,\beta):,
\end{equation}
where $\rho(\beta)$ is a thermal state with inverse temperature $\beta$ and we defined the remaining piece as $:\rho(c,\beta):$,
\begin{align}
:\rho(c,\beta):=\sum_{nm}\mathrm{e}^{-\frac{\beta}{2}(E_n+E_m)}R_{nm}|n\rangle\langle m|
\end{align}
Inspired by the spacetime half-wormhole argument, one may naively illustrate the bulk counterpart of the non-self-averaging part as
\begin{equation}
:\rho(c,\beta):\overset{?}{\simeq}\parbox{\shwl}{\usebox{\boxshwl}}
\end{equation}
Note that the circular direction is just a spatial direction, not a Euclidean time. Based on this naive speculation, the simplest wormhole Wick theorem, which motivated us to call $:\rho(c,\beta):$   ``spatial half-wormhole'', may be illustrated as
\be
\underbrace{\parbox{\shwl}{\usebox{\boxshwl}}}_{:\rho_L(c,\beta):}\otimes\underbrace{\parbox{\shwr}{\usebox{\boxshwr}}}_{:\rho_R(c^*,\beta):}=\underbrace{\parbox{\wh}{\usebox{\boxwh}}}_{\rho^{(\text{TFD})}_{LR}(2\beta)}+:\text{non-self-averaging part}:.
\ee
From this simple calculation (and naive pictorial reason), perhaps it is more natural to call the non-self-averaging part of  single state(or partition function)  ``half-wormhole'' although now the corresponding self-averaging part is the disk contribution, not the wormholes. 

Of course, the above cartoon is speculative. The precise bulk interpretation of $:\rho(c,\beta):$ itself is still unclear. An interesting possibility is that the end of the bulk (fuzzy part of the cartoon) may be identified with the gravity edge modes at the semi-classical level. Recently, a bulk candidate of (spacetime) half-wormholes has been proposed in \cite{Garcia-Garcia:2021squ}. It would be an interesting future direction to find Lorentzian analog of such a bulk geometry, perhaps along the line of \cite{Blommaert:2021etf}.

\subsubsection{Spatial half-wormholes from TPQ}

In \cite{Saad:2021rcu}, they simplified the SYK model by reducing the time direction to one point, where the wormhole can be dominant saddle.
In this section, we consider an analog of this simplification.
As we have shown in the previous subsection, the averaged state has two contributions,
\begin{equation}
\overline{\rho_{LR}(c,\beta)} = \rho_L(\beta) \otimes \rho_R(\beta) + \frac{Z(2\beta)}{Z(\beta)^2} \rho_{LR}^{\text{(TFD)}}(\beta),
\label{eq:avdensity}
\end{equation}
The first term corresponds to a disk contribution and the second corresponds to a wormhole, compared to the SYK model case.
For the comparison with the SYK model with one time point, let us focus on the case where the self-loop contributions (such as $\rho_L$ and $\rho_R$) are negligible. In our example, this is realized in  correlation functions  of operators which are not  associated with conserved charges \footnote{Such a situation is also discussed in \cite{Balasubramanian:2014gla}, for example. }.

In such a correlation function, the expectation value in the canonical ensemble vanishes,
therefore the wormhole contribution can be dominant like the SSSY model. If we strip  off the effective structure of the density matrix \eqref{eq:avdensity} within the correlation function, we obtain
\begin{equation}
\overline{\rho_{LR}(c,\beta)} = \frac{Z(2\beta)}{Z(\beta)^2} \rho_{LR}^{\text{(TFD)}}(\beta).
\end{equation}
Now we are interested in whether there are other contributions in the non-averaged density matrix.
For this purpose, it is useful to see the average of the squared quantity like (\ref{eq:z4}), by preparing a copy of the original system. 
By utilizing the Wick theorem for the random variable $R_{nm}$,
we can show that there are the following contributions in the average,
\footnote{
As we mentioned before, we focus only on conserved quantities.
Therefore, what we are considering is
\begin{equation*}
\overline{
\mathrm{Tr}(\rho_{LR}(c,\beta)\mathcal{O}_L\mathcal{O}_R)
\mathrm{Tr}(\rho_{L'R'}(c,\beta)\mathcal{O}_{L'}\mathcal{O}_{R'})
},
\end{equation*}
and here we abbreviate it to avoid cumbersome expressions.
}
\begin{equation}\label{eq:square}
\overline{\rho_{LR}(c,\beta)^2} = \pa{\frac{Z(2\beta)}{Z(\beta)^2}}^2
\pa{
\rho_{LR}^{\text{(TFD)}}(\beta) \otimes \rho_{L'R'}^{\text{(TFD)}}(\beta)
+\rho_{LR'}^{\text{(TFD)}}(\beta) \otimes \rho_{L'R}^{\text{(TFD)}}(\beta)
+\rho_{LL'}^{\text{(TFD)}}(\beta) \otimes \rho_{RR'}^{\text{(TFD)}}(\beta)
}.
\end{equation}
Here we describe the original systems as $L,R$ and the auxiliary systems as $L',R'$.
This Wick theorem decomposition is completely analogous to (\ref{eq:z4}).
The first term comes from the normal wormhole contribution.
The second and third terms represent the wormhole between the original system and the auxiliary system,
which can be thought of as half-wormhole contributions for the same reason explained below (\ref{eq:normal}). We should emphasize that these results are valid only for non-conserved quantities. Our motivation for this simplification was to discuss situations where wormholes would always be the dominant contributions as in \cite{Saad:2021rcu}. We will generalize our argument to conserved quantities in section \ref{subsec:cons}.

We can naturally expect that the non-averaged state $\rho_{LR}(c,\beta)$ can be approximated by only these two contributions,
wormhole and half-wormhole contributions. This motivates us to consider the following density matrix $\rho$

\begin{equation}\label{eq:bfrho}
\rho =\frac{Z(2\beta)}{Z(\beta)^2}\pa{ \rho_{LR}^{\text{(TFD)}}(\beta)  + :\rho_{LR}(c,\beta):}
\end{equation}
where we abused the normal order symbol for simplicity,
whose precise definition is
\begin{align}
\rho &=
\fr{1}{Z(\beta)^2}
\sum_{n,m,a,b}\mathrm{e}^{-\frac{\beta}{2}(E_n+E_m+E_a+E_b)} \overline{R_{nm} R_{ba}} \ket{n_La_R}\hspace{-1mm}\bra{m_Lb_R} \\ \nonumber
&+
\fr{1}{Z(\beta)^2}
\sum_{n,m,a,b}\mathrm{e}^{-\frac{\beta}{2}(E_n+E_m+E_a+E_b)} :R_{nm} R_{ba}: \ket{n_La_R}\hspace{-1mm}\bra{m_Lb_R}.
\end{align}
This is completely analogous to (\ref{eq:normal}) in the SSSY model (where $ \mathbf{\rho}$ corresponds to $\mathbf{z}^2$). It is worth stressing that the inclusion of $:\rho_{LR}(c,\beta):$ recovers the factorized result for the expectation value before the averaging.

To justify that this is a good approximation, we need to estimate the error (like $\braket{ (z^2 - \mathbf{z}^2)^2  }$ in the SSSY model).
Let us consider the error,
\begin{equation}
(\text{Error}) = \rho_{LR}(c,\beta) -\mathbf{\rho},
\end{equation}
then we can show by a brute calculation,
\begin{equation}
\overline{(\text{Error})^2} = \overline{(\rho_{LR}(c,\beta) -\mathbf{\rho})^2} = 0.
\end{equation}
Therefore, we can conclude that the product state (\ref{eq:product}) has a very similar structure as found in the SSSY model. Note that unlike in the SSSY model, the effective state (\ref{eq:bfrho}) is not just a good approximation but an exact expression. 


There is an interesting subsequent work \cite{Mukhametzhanov:2021nea}, which states that the half-wormhole configuration automatically appears as an excitation of the wormhole, like the result \cite{Eberhardt:2021jvj}.
It would be interesting to address this issue in our model.

\subsubsection{Spatial half-wormholes from ETH}

Here, we will comment on the ETH viewpoint. 
As one can expect, the same structure can be found in the ETH. 
For simplicity, we again restrict observables to charged operators. 
Under these assumptions, in the same way as (\ref{eq:square}), the ETH averaging of the squared quantity leads to the same half-wormhole contributions.
Thus, we can conclude that the state $\rho_{LR}(\beta)$ with ETH also has two contributions, wormhole contribution and half-wormhole contribution.

\subsection{Generalization to conserved quantities}\label{subsec:cons}

We have discussed the random product states and their connection to ``spatial half-wormholes'' contribution, especially for non-conserved quantities. In the rest of this section, we generalize it to conserved quantities. 

In particular, the contributions up to sub-sub leading order schematically become
\begin{align}
&\overline{\bra{\Psi_\beta}\mathcal{O}_L\mathcal{O}_R\ket{\Psi_\beta}\bra{\Psi_\beta}\mathcal{O}_{L^\prime}\mathcal{O}_{R^\prime}\ket{\Psi_\beta}}\nn\\
&\simeq \braket{\mathcal{O}}_\beta^4+6\frac{Z(2\beta)}{Z(\beta)^2}\braket{\mathcal{O}}_\beta^2\braket{\text{TFD}_{2\beta}|\mathcal{O}\mathcal{O}|\text{TFD}_{2\beta}}+3\frac{Z(2\beta)^2}{Z(\beta)^4}\braket{\text{TFD}_{2\beta}|\mathcal{O}\mathcal{O}|\text{TFD}_{2\beta}}^2, \label{eq:4pt}
\end{align}
where we suppressed the subscript of subsystems for simplicity. Each numerical coefficient is easily understood as the number of possible  Wick contractions which lead to the term, when we ignore the each label of subsystems. Interestingly, further higher order terms (not written here) can be understood as a generalization of the bra-ket wormholes\cite{PhysRevD.34.2267, Chen:2020tes}. We discuss this aspect in the next section. 
It is also worth noting that the aforementioned higher order terms do not show up if we use the averaging over matrix elements as ETH. As already mentioned in section \ref{sec:ETH}, this difference originates from the difference between the random vector $c_n$ and the random matrix $R_{nm}$\footnote{For example, $\overline{R_{ab}R_{cd}R_{ef}}$ does vanish in ETH, while not in TPQ states. For the definition of $R_{ab}$ in TPQ states, see equation \eqref{eq:RTPQ}.}. 

Notice that the second term of \eqref{eq:4pt} has only one wormhole connection. This contribution is more dominant than the two wormholes contribution. Since the ordinary SYK model should have a non-vanishing one-point function, $\braket{z}\neq0$, we expect that a similar contribution should also appear if we generalize the results of the SSSY model to the ordinary SYK model. 

\section{Discussion}\label{sec:discussion}
We have argued that the contribution of the spatial wormhole emerges when the classically correlated product state is coarse-grained. In particular, we have shown that the spatial analog of half-wormholes is a crucial contribution to solving the factorization puzzle caused by this wormhole. In the remainder of this paper, we discuss the ambiguities coming from choice of measure for the state averaging ({\it i.e.} coarse-graining) and a path-integral interpretation of our wormholes as a consequence of state averaging.
\subsection{Ambiguities from choice of measure} 
In this section, we discuss ambiguities coming from the choice of the measure for averaging. In this paper, we have been studying  the product of the TPQ states,
\be
\ket{\Psi_\beta}=\ket{\psi_{\beta L}}\ket{\psi_{\beta R}^\ast}=\sum_{n,m}\ex{-\frac{\beta}{2}(E_n+E_m)}c_nc_m^*\ket{n_L}\ket{m_R},
\ee
in the presence of  an ensemble average for random coefficients $c_n$ over the Gaussian distribution. The important feature of this prescription is that the ensemble average changes the norm of the product state;
\begin{align}
    &\overline{\langle\Psi_\beta |\Psi_\beta\rangle}=1+\frac{Z(2\beta)}{Z(\beta)^2}\, ,\\   &\overline{\langle\Psi_\beta |\Psi_\beta\rangle^2}=1+6\frac{Z(2\beta)}{Z(\beta)^2}+3\frac{Z(2\beta)^2}{Z(\beta)^4}+8\frac{Z(3\beta)}{Z(\beta)^3}+6\frac{Z(4\beta)}{Z(\beta)^4}\, ,
    \end{align}
    e.t.c.
    The normalization affects, for example $n$-th R\'enyi entropy between the left and the right CFT \footnote{This should not be confused by the results in subsection \ref{subsec:ee}. In subsection \ref{subsec:ee}, we evaluated the entanglement entropy from the averaged density matrix $\bar{\rho}_{LR}$ (\ref{eq:CS}) without using the replica trick, and obtained a physically reasonable result.}
 \be
 S^{(n)}_{LR}=-\frac{1}{1-n}\log \overline{{\rm tr}\rho_R^n}\, ,
\ee
where $\overline{{\rm tr}\rho_R^n}$ is defined by
\begin{align}
\overline{{\rm tr}\rho_R^n}=\overline{\langle\Psi_\beta |\Psi_\beta\rangle^n}/\left(\overline{\langle\Psi_\beta |\Psi_\beta\rangle}\right)^n\, .
\end{align}
Notice that the ensemble average is taken for the numerator and the denominator (normalization) separately. This prescription is often used when we compute the entanglement entropy in gravitational systems by the replica trick (for example, see \cite{Penington:2019kki} ). However, as we will see below, this gives an unphysical value of the R\'enyi entropy in our case. 

To see this, we first point out  that each way of Wick-contractions between $c_n$ always gives a positive contribution. As a result, one can easily show that 
\begin{align}\overline{\langle\Psi_\beta |\Psi_\beta\rangle^n}>\left(\overline{\langle\Psi_\beta |\Psi_\beta\rangle}\right)^n\, ,
\end{align}
for $n>1$, which leads to the negative R\'enyi entropy. As we already saw, this unphysical feature of the R\'enyi entropy is rooted in our prescription where we take the ensemble average of the numerator and the normalization separately.

To avoid this unphysical result, one can instead use a different measure of the ensemble averaging that preserves the normalization of the state. Here we will give a brief introduction of one of such measures. We introduce the appropriate measure which  (i) reproduces the canonical thermal correlators from random states $|\psi\rangle=\sum_{n}\psi_n|n\rangle$\footnote{Notice that the coefficients of $|\psi\rangle$ do not include the Boltzmann factors $p_n$. The corresponding factors in the thermal correlators computed by averaging these state come from the measure itself.} and (i) preserves the norm of the state $\|\psi\|=\langle\psi|\psi\rangle$. One example of such measures is called the Gaussian adjusted projected (GAP) measure\cite{Goldstein_2006}, which is also introduced in \cite{Freivogel:2021ivu}. The GAP measure is explicitly written as
\begin{align}
D \psi=\delta(1-\|\psi\|)\left(\prod_{l=1}^{d} \frac{d \psi_{l}^{*} d \psi_{l}}{2 \pi i} \frac{Z(\beta)}{p_{l}}\right) \int_{0}^{\infty} d r r^{2 d+1} \exp \left(-r^{2} \sum_{n} \frac{Z(\beta)}{p_{n}}\left|\psi_{n}\right|^{2}\right)\, .
\end{align}
As we can see, the delta function $\delta(1-\|\psi\|)$ in the measure projects onto the appropriately normalized averaged states, i.e, $\overline{\|\psi\|^n}=\overline{\langle\psi|\psi\rangle^n}=1$. Since we have $\langle\Psi_\beta |\Psi_\beta\rangle^n=\langle\psi|\psi\rangle^{2n}$, the trace of the $n$-th power of the density matrix  $\overline{{\rm tr}\rho_R^n}$ is normalized to be unity
\begin{align}
\overline{{\rm tr}\rho_R^n}=\overline{\langle\Psi_\beta |\Psi_\beta\rangle^n}=1\, ,
\end{align}
which gives the trivial entanglement entropy. This is rooted in the fact that our original fine-grained
state is a product state that has zero entanglement entropy.

The averaged two-point function of  the random coefficients $\psi_n$ is given by
\begin{align}
\overline{\psi_{n} \psi_{m}^{*}}=\frac{p_{n}}{Z(\beta)} \delta_{n m}
\end{align}
 with $p_n=e^{-\beta E_n}$, which reproduces $ \bar{\rho}=\sum_{n, m} \overline{\psi_{n}^{*} \psi_{m}}|n\rangle\langle m|= \sum_{n} p_{n} |n \rangle \langle n| /Z(\beta)$, {\it i.e.}, the density matrix in the canonical ensemble as desired.  Importantly, this measure changes the averaged two-point functions for the product state
 \be
 |\Psi\rangle=|\psi_L\rangle|\psi^*_R\rangle\, ,
 \ee
which we already computed in \eqref{eq:2pt0} with our original measure.  This is computed from the four-point function of the random coefficients $\psi_n$, which is explicitly given by 
\begin{align}\label{eq:four-point}
\overline{\psi_{n_{1}} \psi_{n_{2}} \psi_{m_{1}}^{*} \psi_{m_{2}}^{*}} \simeq\frac{p_{n_{1}} p_{n_{2}}}{Z(\beta)^{2}}\left(\delta_{n_{1} m_{1}} \delta_{n_{2} m_{2}}+\delta_{n_{1} m_{2}} \delta_{n_{2} m_{1}}\right)\left(1+\frac{Z(2\beta)}{Z(\beta)^{2}}-\frac{p_{n_{1}}}{Z(\beta)}-\frac{p_{n_{2}}}{Z(\beta)}\right)\,  .
\end{align}
Here we take the thermodynamic limit of 2D CFTs where $Z(\beta)\sim e^{S/2}$, and omit $O(e^{-S})$ terms. From this, the averaged two-point function for the product state $|\Psi\rangle$ is computed as 
\begin{align}\label{eq:GAPtwo}
&\overline{\braket{\Psi|\mathcal{O}_L\mathcal{O}_R|\Psi}}\nn\\
&\simeq \braket{\mathcal{O}_L}_{\beta}\braket{\mathcal{O}_R}_{\beta}+\frac{Z(2\beta)}{Z(\beta)^2}\braket{\text{TFD}_{2\beta}|(\mathcal{O}_L-\braket{\mathcal{O}_L}_{\beta})(\mathcal{O}_R-\braket{\mathcal{O}_R}_{\beta})|\text{TFD}_{2\beta}}. 
\end{align}
Compared with \eqref{eq:2pt0}, we have additional negative contributions. They are necessary since when ${\cal O}_{L,R}=\mathbbm{1}$, the second term in \eqref{eq:GAPtwo}  should vanish due to the normalization $\overline{\braket{\Psi|\Psi}}\nn=1$.

\subsection{Relation to the spacetime wormholes in the gravitational path-integral}
\begin{figure}[t]
\centering
\includegraphics[scale=0.3]{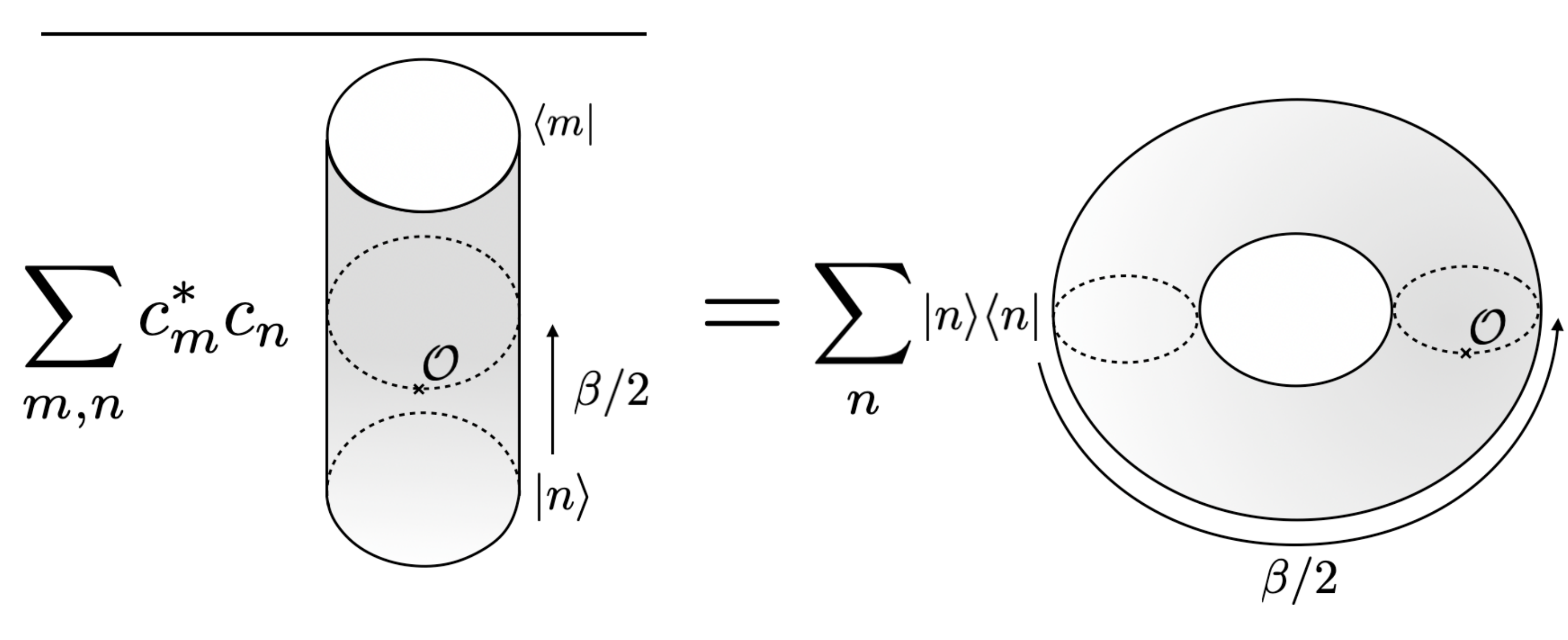}
\caption{The geometrical picture of how the canonical TPQ state reproduces the thermal expectation values. This has a similar geometrical structure to the bra-ket wormhole in the gravitational path-integral.\vspace{3mm}}\label{fig:TPQ2}
\includegraphics[scale=0.20]{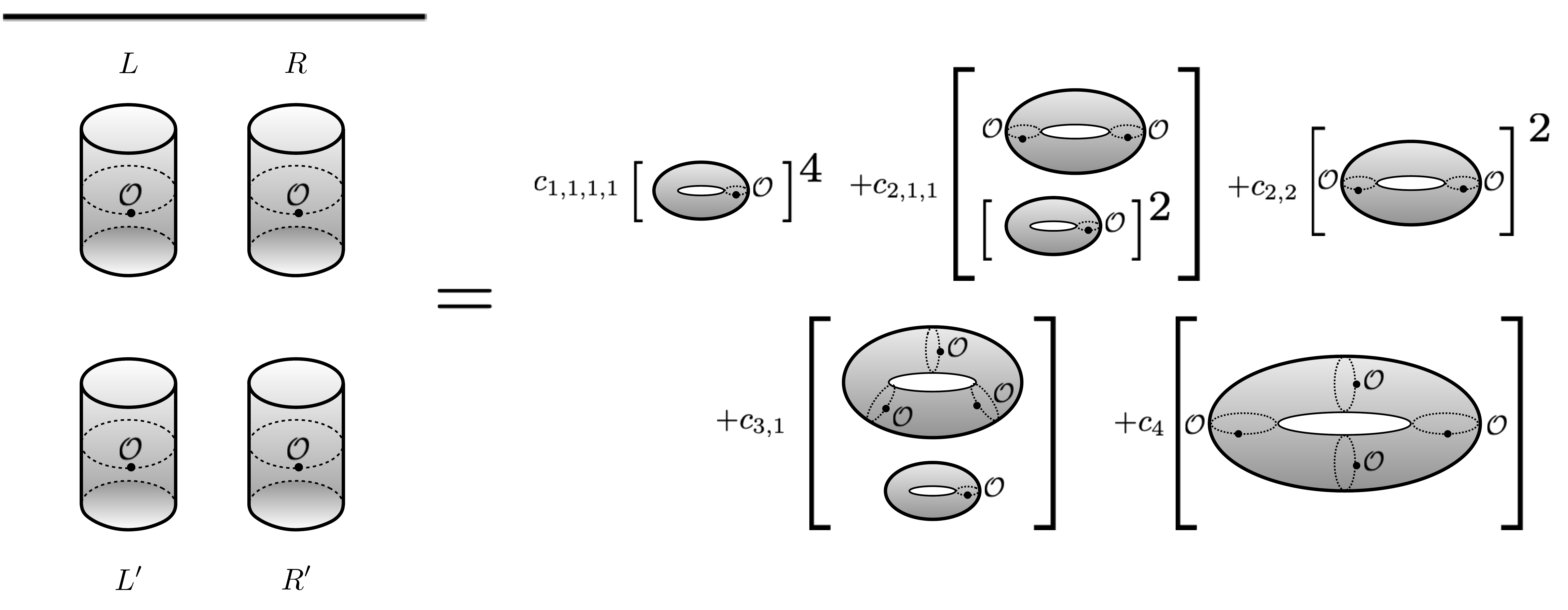}
\caption{A schematic picture for higher moments calculation in TPQ states as \eqref{eq:4pt}. In particular, we can see the generalization of the bra-ket wormholes. (In \eqref{eq:4pt}, we neglected the forth and fifth terms as these are further suppressed by the power of $e^{-S}$.) Note that we do not describe the detail of each operator. Also, we represent all operators in terms of a single CFT.}\label{fig:4pt}
\end{figure}
In this paper, we studied emergence of the {\it spatial} wormholes from averaging over classically-correlated states. We found similar geometric structures to the {\it spacetime} wormholes which appear after averaging over theories. One apparent difference between these two types of wormholes is that a spacetime wormhole appears in the gravitational path-integral with a fixed topology of the boundary manifold, while emergence of a spatial wormhole studied in this paper changes the topology of its boundary manifold itself (see Fig.\ref{fig:TPQ2} and Fig.\ref{fig:4pt}). Therefore one might consider that emergence of the spatial wormholes studied in this paper has no relevance to that of the spacetime wormholes in the gravitational path-integral, which plays key roles in understanding holography\cite{Saad:2018bqo} and the black hole information paradox\cite{Penington:2019kki, Almheiri:2019qdq}. However, in a recent paper\cite{Chen:2020tes}, it was argued that one should also take account of geometries with different topologies of boundary manifolds in the gravitational path-integral to avoid so-called SSA (= Strong Sub-Additivity) paradox. As a consequence, one needs to include a new geometry called {\it bra-ket} wormhole that connects the bra with the ket of the original manifold. In this paper, we saw the canonical TPQ states reproduce the thermal expectation value in section \ref{sec:TPQ}. In this case, averaging over random coefficients in the expression of the TPQ state (equation  (\ref{eq:cTPQ})) leads to connection between the bra and the ket and that creates a thermal circle to reproduce the thermal behavior of the correlators (see Fig.\ref{fig:TPQ2}). This has a similar geometrical structure to the bra-ket wormhole in the gravitational path-integral.
From this view point, the wormholes that appear after averaging over classically-correlated product states can be viewed as generalizations of the bra-ket wormhole since they are created by connecting the bras and the kets of the product states respectively (see Fig.\ref{fig:4pt}). 
It would be interesting to understanding emergence of these wormholes in terms of gravitational path-integral in future work.

\section*{Acknowledgement}
We thank Yasunori Nomura, Masahiro Nozaki, Kenta Suzuki, Tadashi Takayanagi, and Zixia Wei for useful discussion. KG is supported by JSPS Grant-in-Aid for Early-Career Scientists 21K13930. KT is supported by JSPS Grant-in-Aid for Early-Career Scientists 21K13920. TU is supported by JSPS Grant-in-Aid for Young Scientists 19K14716. YK is supported by RIKEN iTHEMS Program, and the RIKEN Special Postdoctoral Researcher program.

\bibliographystyle{JHEP}
\bibliography{main.bib}

\end{document}